\documentclass[preprint,prl,aps,onecolumn,preprintnumbers,amsmath,
amssymb,amsfonts,letterpaper]{revtex4-1}
\usepackage{graphicx,amsmath}
\usepackage{dcolumn}
\begin{document}

\title{Metastability in pressure-induced structural transformations of CdSe/ZnS
core/shell
nanocrystals}

\author{Michael Gr\"unwald}
\affiliation{Department of Chemistry,
  University of California, Berkeley, California 94720}

\author{Katie Lutker}
\affiliation{Department of Chemistry, University
  of California, Berkeley, California 94720}

\author{A. Paul Alivisatos}
\affiliation{Department of Chemistry,
  University of California, Berkeley, and Materials Sciences Division,
  Lawrence Berkeley National Laboratories, Berkeley, California
  94720}

\author{Eran Rabani}
\affiliation{School of Chemistry, The Sackler
  Faculty of Exact Sciences, Tel Aviv University, Tel Aviv 69978,
  Israel}

\author{Phillip L. Geissler}
\affiliation{Department of Chemistry,
  University of California, Berkeley, and
  Lawrence Berkeley National Laboratories, Berkeley, California
  94720\,}

\begin{abstract}
\bf
The kinetics and thermodynamics of structural transformations under
pressure depend strongly on particle size due to the influence of
surface free
energy \cite{Alivisatos1994,Alivisatos1997,Alivisatos2001,Grunwald2009}. By
suitable design of surface structure \cite{Grunwald2006}, composition \cite{Farvid2009}
and passivation \cite{Alivisatos1997} it is possible, in principle, to prepare nanocrystals in
structures inaccessible to bulk materials \cite{Alivisatos2002a}. 
However, few realizations of such extreme size-dependent behavior exist
\cite{Bawendi1999}. 
Here we show with molecular dynamics 
computer simulation that in a model of CdSe/ZnS core/shell
nanocrystals the core high pressure structure can be made metastable under
ambient conditions by tuning the thickness of the shell. In nanocrystals
with thick shells, we furthermore observe a wurtzite to NiAs 
transformation, which does not occur in the pure bulk materials. These
phenomena are linked to a fundamental change in the atomistic
transformation mechanism from heterogenous nucleation at the surface to
homogenous nucleation in the crystal core. Our results suggest a new
route towards expanding the range of available nanoscale materials.
\end{abstract}

\maketitle

At thermodynamic equilibrium, matter adopts the form
that minimizes its total free energy~\cite{Callen1985}. Close to
first-order phase transitions, however, metastability of the competing
phases is 
often
observed; liquid water can be cooled many degrees
below its freezing point, magnets can withstand oppositely oriented
magnetic fields, and diamonds do not transform to graphite
at ambient conditions. 
How far
one can push a system out of its equilibrium phase depends on the
microscopic transformation mechanism that determines the height of the
free energy barrier separating the competing phases.

Exploiting the metastability of different solid phases is a possible
route to creating materials with new properties. However, many crystal
structures form only when high pressure is applied and are unstable
under ambient conditions in the bulk. In nanocrystals, on the other
hand, phase diagrams and microscopic transformation mechanisms can
depend strongly on 
particles' 
size and shape~\cite{Grunwald2006,Bealing2010a}. The
wurtzite to rocksalt transformation in CdSe nanocrystals, for example,
shows an increasing thermodynamic transition pressure and a decreasing
activation enthalpy with decreasing particles
size~\cite{Alivisatos1995,Alivisatos1997,Grunwald2009a}. While it is
in principle possible to extend the metastability of high-pressure
structures to ambient conditions by engineering the surface properties
of nanoparticles, significant insight into the underlying microscopic
transformation dynamics is required.

A particularly interesting surface modification is realized in core/shell
nanocrystals~\cite{Spanhel1987}, where the core material is
epitaxially overgrown with a material of identical crystal
structure~\cite{Hines1996,Bawendi1997,Peng1997}. While the optical qualities of
these hetero-materials are
well-studied~\cite{Rosenthal2007}, little is known about their
structural properties~\cite{Li2011}. In modern synthesis methods,
materials with a lattice mismatch of up to $11\%$ can be combined to
form a pristine core/shell interface~\cite{Rosenthal2007}.
The resulting lattices of both
core and shell experience a strong strain that depends sensitively on
particle size and has the potential of introducing dramatic changes to
the nanoparticle's structural and kinetic behavior under pressure.

In this Letter, we report the simulation of spherical wurtzite CdSe
nanocrystals of $3$~nm diameter ($510$~atoms), that have been
epitaxially passivated with ZnS shells of thicknesses up to $2.1$~nm
($5$~monolayers). The largest of these core/shell crystals consists of
$6918$~atoms. The particles are modeled with empirical pair potentials
designed to reproduce a number of properties of the bulk
materials~\cite{Rabani2001,Rabani2002}. In our simulations,
a single crystal is immersed in a pressure bath of ideal gas
particles~\cite{Grunwald2006a,Grunwald2007} at a temperature of
$300$~K and the pressure is increased in steps of $0.2$~GPa every
$10$~ps. These pressurization rates are many orders of magnitude
larger than in experiments using diamond anvil cells but are
comparable to recent shock-wave experiments on CdSe
nanocrystals~\cite{Wittenberg2009}. When a pressure of $20$~GPa is
reached after $1$~ns, the pressure is released again at the same
rate. After reaching ambient pressures, the crystals are simulated for
another $1.2$~ns. We monitor the evolution of the crystal structure by
calculating atom-coordination numbers based on the radial pair
distribution functions of core and shell atoms.

The effect of the ZnS shell on the structure of the CdSe core is
dramatic. In Figure~\ref{dens_fig}B we plot the density of the
wurtzite core of crystals with different shell sizes as a function of
external pressure. The density of the core increases significantly
with increasing
shell thickness. For a $2$~nm shell, this
compression effect is equivalent to an additional external pressure of
6 GPa, as illustrated in Figure \ref{dens_fig}C. This pressure is much
higher than the bulk coexistence pressure of $2.4$~GPa of our CdSe model, and high
enough to cause spontaneous transformation in bare CdSe crystals. Similarly high pressures
were found at the core/shell interface in experiments of CdS/ZnS nanocrystals \cite{Ithurria2007}.
One might therefore expect the transformation in core/shell crystals to happen at
lower pressures compared to bare CdSe nanocrystals. Quite to the
contrary, the upstroke transformation pressure of the core increases
strongly with increasing shell size, as illustrated in
Figure~\ref{transp_fig}. While pure CdSe nanocrystals transform at
around $6$~GPa, transformation pressures of up to $18$~GPa are
observed for crystals with thick shells.

Our simulations suggest that the increase in upstroke transformation pressure
with particle size is caused by an increase in thermodynamic
transition pressure and a concurrent removal of favorable surface
nucleation sites. We estimated the phase diagram of the nanoparticles
by calculating the pressures at which crystals in different phases
have equal enthalpy (see Figure~\ref{transp_fig}). In particular, we
consider three combinations of core/shell crystal structures:
wurtzite/wurtzite, rocksalt/wurtzite, and rocksalt/rocksalt. A shell
thickness of $0.5$~nm, corresponding to a single monolayer of ZnS, is
enough to raise the
thermodynamic transition pressure 
by $2$~GPa. With increasing
shell size, the phase boundary of the all-rocksalt phase approaches
the bulk thermodynamic transition pressure of our ZnS model. At larger shell
sizes ($3$--$4$ monolayers), the rocksalt/wurtzite phase, featuring distinct
crystal structures in the core and shell, appears as a stable
intermediate between the homogenous phases. The thermodynamic transformation
pressure from the all-wurtzite to the heterogenous phase is fairly
insensitive to shell thickness, indicating that a large-shell regime
has been reached. 

The appearance of the rocksalt/wurtzite phase is accompanied by a
fundamental change of nucleation mechanism. Computer simulations of
pure CdSe nanocrystals have shown that crystals with low energy
surface facets always transform via nucleation events on the
surface.\cite{Grunwald2009,Grunwald2009a} While we observe similar
surface nucleation in core/shell crystals with shell thicknesses up to $3$
monolayers, for crystals with thick shells nucleation happens in the core,
as illustrated in
Figure~\ref{nucl_fig}. Despite the nearly constant thermodynamic
transition pressure in this size-regime, the observed transformation
pressures still increased.  This indicates that the change from
heterogeneous to homogeneous nucleation results in an increase in the
size of the nucleation barrier.

While a typical core transformation event lasts no longer than
$10$~ps, transformations of shells proceed in steps, creating only
confined regions of rocksalt at a time. 
The hysteresis curves in Figure~\ref{nucl_fig}
manifest such dynamics.
By $20$~GPa, however, most shells have completed the transformation to rocksalt
By comparison, we found that a pure $4$~nm ZnS nanocrystal
remained in the wurtzite structure when subjected to the same pressure
protocol, indicating that the core also influences the shell. 
Upon release of pressure, all crystal shells undergo a
back-transformation: Thick shells transform back to a mixture of
wurtzite/zinc-blende, thin shells to predominantly amorphous
four-coordinated structures.

Not all crystal cores, on the other hand, underwent a back-transformation.
While crystal cores with thin shells ($\lesssim 1$~monolayers) and thick shells ($\gtrsim 4$ ~monolayers) transform back to mixtures of wurtzite and zinc-blende structures, cores in a broad range of intermediate
shell sizes remain in the rocksalt structure down to zero pressure,
as illustrated in Figure~\ref{meta_fig}. This observation suggests a significant increase of the down-stroke kinetic barrier with increasing shell thickness. The nature of this increased barrier height is related to the interface between the rocksalt core and the re-transformed shell. During the back-transformation of the shell, bonds between core and shell atoms stay mostly intact. The re-transformed shell neatly passivates the rocksalt core and thus removes favorable nucleation sites at the core surface. The amorphization of shells thinner than 1~nm upon back-transformation is a consequence of this passivation and the mismatch between the rocksalt and wurtzite structures in the core and shell, respectively. Very thick shells, however, transform back into crystalline wurtzite/zinc-blende mixtures and induce larger stress at the core/shell interface, facilitating the back-transformation of the core. Our simulations demonstrate that rocksalt cores can persist on the nanosecond timescale; recent experiments suggest substantial
metastability even on experimental timescales~\cite{Li2011}. 

Phase transitions that occur far from equilibrium do not necessarily
lead to the phase with the lowest free energy. In fact, Ostwald's step
rule predicts that a system will transform from a metastable phase to the phase
with the smallest free energy difference.
In an unexpected realization of this
rule of thumb, we observed a wurtzite to NiAs (B8) transformation in a
few crystals with thick shells. Figure~\ref{NiAs_fig}A shows three
snapshots of a $1.9$~nm shell crystal. In the course of the
transformation a grain boundary between the expected rocksalt
structure and the NiAs structure builds up in the core and later
propagates into the shell at higher pressures. Like rocksalt, the NiAs
structure is $6$-coordinated. Cations are in a rocksalt-type
coordination environment, while anions are coordinated by a trigonal
prism of cations (Figure~\ref{NiAs_fig}B). The occurrence of the NiAs
structure is surprising, since it has not been observed experimentally
in the pure materials, neither in the bulk nor in
nanocrystals. Figure~\ref{NiAs_fig}C shows a plot of the bulk
enthalpies of the core and shell materials in the wurtzite, rocksalt
and NiAs structures as a function of pressure. Throughout the pressure
range studied here ($0$--$20$ GPa), NiAs is never enthalpically
most
stable. However, it becomes metastable with respect to the wurtzite
structure at pressures larger than $4.5$~GPa and $16$~GPa, for CdSe and
ZnS, respectively. (The transformation illustrated in Figure
\ref{NiAs_fig}C occurred at 17 GPa.)  Interestingly, in a recent
pressure study of ZnS/CdSe core/shell nanocrystals an unexpected
Raman peak was observed after the transformation had
happened~\cite{Fan2007}. 

In summary, we have shown that both the kinetics and thermodynamics of the
wurtzite to rocksalt transformation in CdSe/ZnS core/shell crystals are strongly
affected by the thickness of the shell. A strong increase in thermodynamic
transition pressure with increasing shell thickness is accompanied by a
substantial broadening of the hysteresis, rendering the transformed rocksalt
cores metastable at ambient conditions. The up-stroke nucleation pathway changes
from heterogenous nucleation on the surface to homogenous nucleation in the
core. In thick-shell crystals, the greatly increased up-stroke transformation
pressure can lead to nucleation of the NiAs structure, which is not observed in
the two pure materials.

The unexpected occurrence of a new high-pressure NiAs structure suggests that other
materials might be susceptible to a similar phenomenon. By
artificially increasing the pressures at which solid-solid
transformation take place, transformation routes to other, previously
unobservable crystal structures might become available. Potentially,
such an increase can be achieved by blocking favorable nucleation pathways
through suitable surface modifications, or by using high pressurization rates
as obtained in shockwave experiments.

\subsection{Methods}

The interatomic pair potential for ZnS was obtained using the same procedure outlined in Ref.~\cite{Rabani2002} and consists of a Lennard-Jones part and electrostatic interactions. Potential parameters are given in Table \ref{paratable}. Standard mixing rules were used for interactions between unlike atom types. The equations of motion were integrated with the velocity Verlet algorithm \cite{Frenkel2002} and a time step of $2$~fs. The ideal gas barostat was parameterized as in Ref.~\cite{Grunwald2007}. 

\begin{table}[h]
\begin{center}
\begin{tabular*}{50mm}{@{\extracolsep{\fill}}c|ccc}
& $q(\mathrm{e})$ & $\sigma \mbox{(\AA)}$ & $\epsilon \mbox{(K)}$ \\\hline
Zn & 1.18 & 0.02 & 17998.4  \\
S & -1.18 & 4.90 & 16.5  \\
Cd & 1.18 & 1.98 & 16.8  \\
Se & -1.18 & 5.24 & 14.9 \\
\end{tabular*}
\end{center}
\caption{\label{paratable} Potential parameters for ZnS and CdSe \cite{Rabani2002}.}
\label{ta:par}
\end{table}

To generate core/shell nanocrystals, we place Cd, Se, Zn, and S atoms
on a contiguous subset of the lattice sites that define a bulk
wurtzite crystal with lattice constants $a=4.3$~\AA{} and
$c=7.0$~\AA{}. A roughly spherical shape is obtained by including in this subset only lattice
sites lying within a distance $r=15\,\mathrm{\AA}+r_\mathrm{shell}$
away from the center of a particular (arbitrary) unit cell. Surface
energies are reduced by subsequently excluding lattice sites
possessing fewer than two nearest neighbors. Atoms are positioned on
the resulting subset of lattice sites as follows: (i) Cd and Se are
placed on all ``a'' and ``b'' sites, respectively, within a radius of
$15$~\AA{} of the center.  (ii) Zn and S atoms are placed on all
remaining ``a'' and ``b'' sites, respectively. (iii) In a final step
designed to mimic effects of CdSe core relaxation prior to shell
growth, all Cd and Se atoms bound to fewer than two Se and Cd atoms,
respectively, are replaced by Zn and S atoms, respectively.

\begin{figure}
\includegraphics[width=0.4\textwidth]{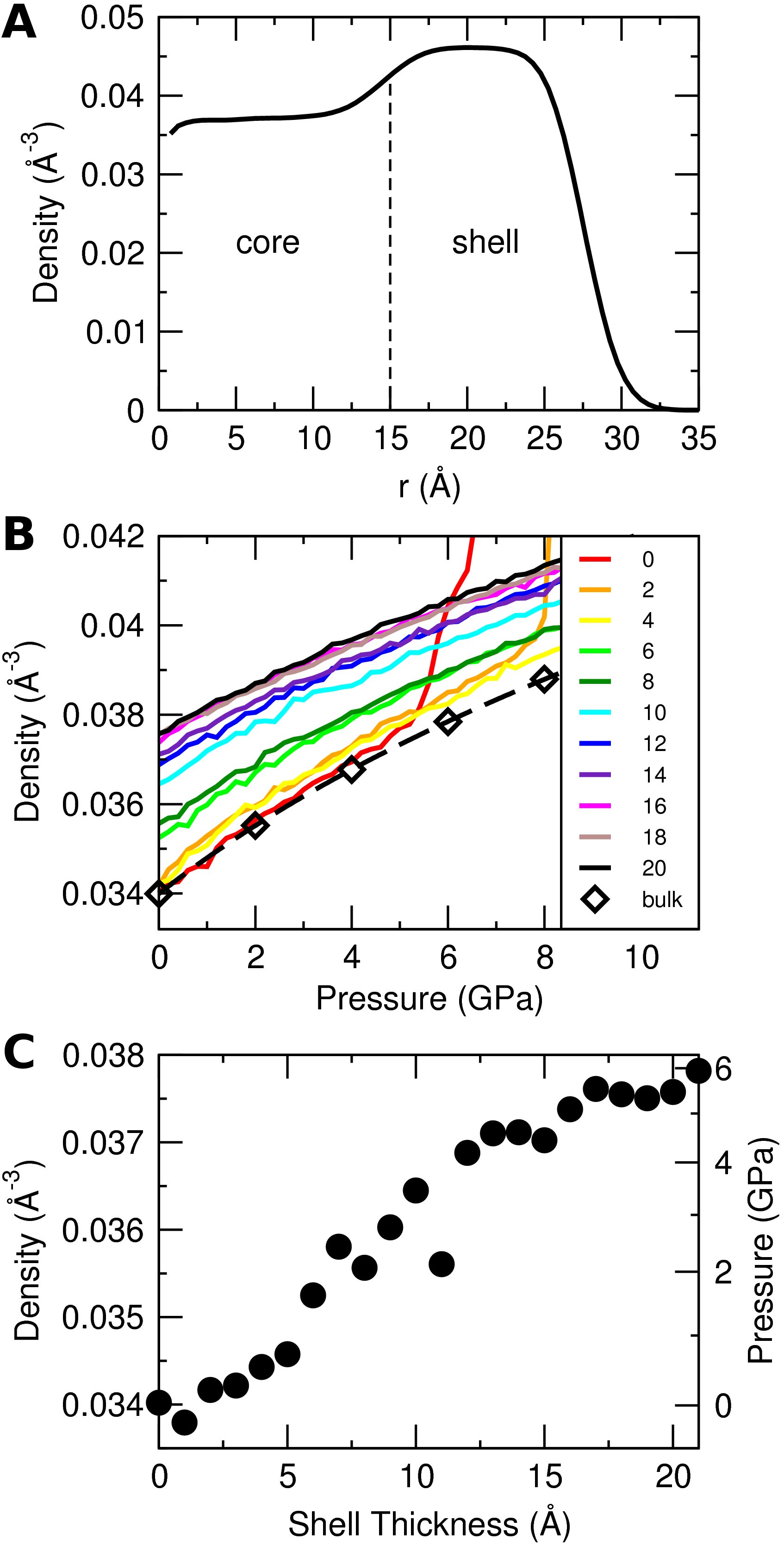}
\caption{\label{dens_fig} {\bf The ZnS shell compresses the CdSe core.}
 ({\bf A}) Coarse-grained atom density 
of a $1.5$~nm shell-crystal at
zero pressure as a function of distance $r$ from the center of mass. The
different densities of the core and shell materials are well
visible. ({\bf B}) Density of the core as a function of pressure,
  for nanocrystals with different shell thickness (legend values
  indicate shell thickness in \AA{}). The density of bulk CdSe,
  obtained from constant pressure Monte Carlo simulations, is shown
  for reference.  The dashed line is a fit of the bulk data to the
  Murnaghan equation of state.  Note that the sudden increase in
  density observed for $0$ and $0.2$~nm shell-crystals is a signature
  of the wurtzite to rocksalt transformation. ({\bf C}) Core density
  at zero pressure as a function of shell thickness. The right hand
  ordinate shows the pressure necessary to achieve 
equivalent densities in bulk CdSe.}
\end{figure}

Coarse-grained atom densities were computed as
\begin{equation}\label{cgdens}
\bar{\rho}(\mathbf{r}) = \left(2\pi\sigma^2 \right)^{-3/2} \sum_{i=1}^N \exp\left(\frac{(\mathbf{r}-\mathbf{r}_i)^2}{2\sigma^2}\right),
\end{equation}
where 
$\mathbf{r}_i$ is the position of atom $i$, 
and $N$ is the total number of atoms.
The effective atomic width
$\sigma=2$~\AA{} 
was chosen to
balance
smoothness and resolution of $\bar{\rho}(\mathbf{r})$.
The density profile $\rho(r)$ plotted in 
Figure \ref{dens_fig}A 
was computed by averaging $\bar{\rho}(\mathbf{r})$ within thin spherical
shells centered on the nanocrystal's center of mass:
The core density
values in Figure \ref{dens_fig}B
were further averaged over distances within 
$8$~\AA{} of the nanocrystal's center of mass.

\begin{figure}
\includegraphics[width=0.6\textwidth]{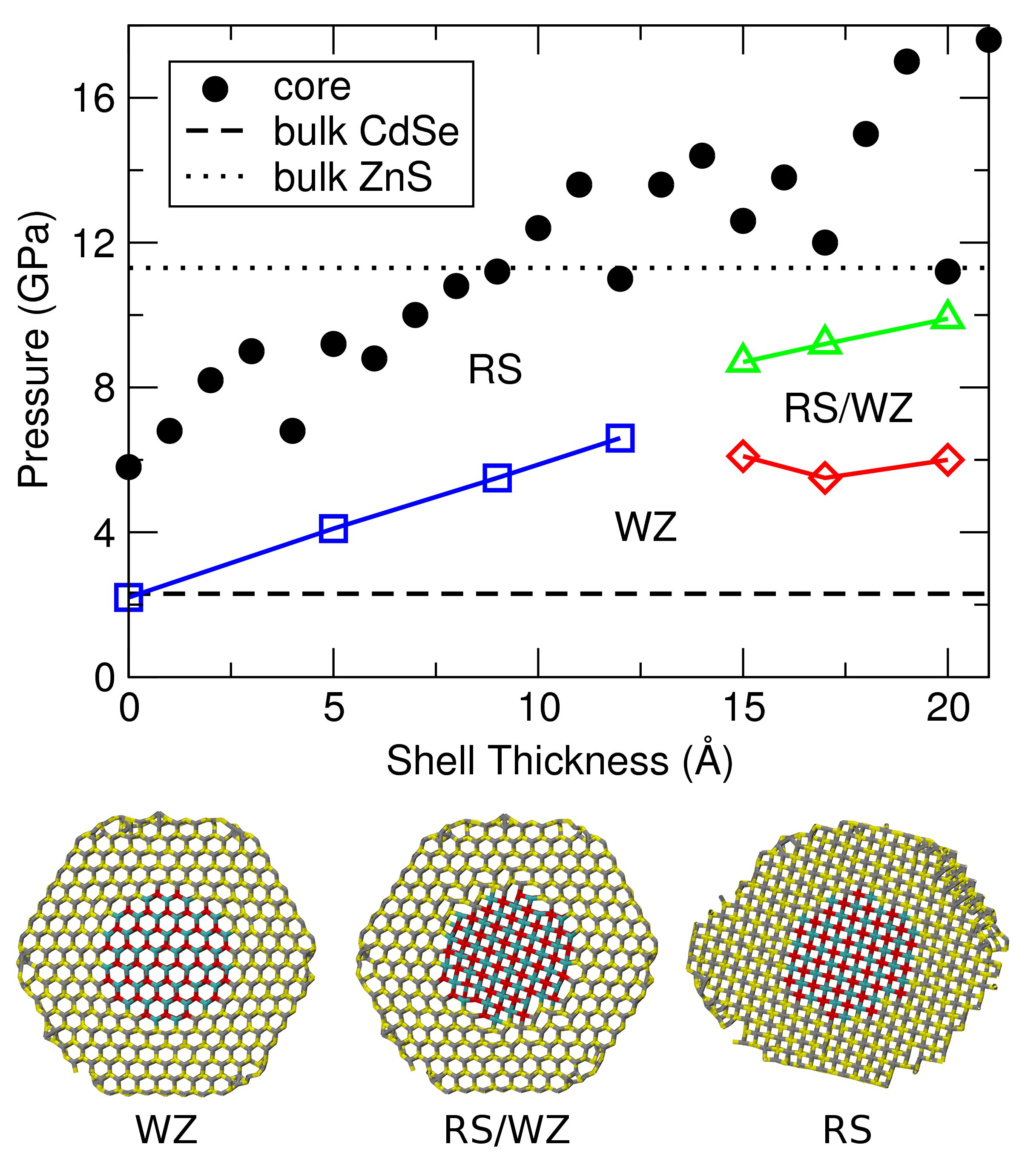}
\caption{\label{transp_fig} {\bf Size-dependent transformation
    pressure.} Core upstroke transformation pressures (solid black
  circles) are plotted as a function of shell thickness. At these
  pressures, the fraction of six-coordinated atoms exceeds $0.1$ for
  the first time. The thermodynamic transition pressure of bulk CdSe
  (dashed line) and bulk ZnS (dotted line) are shown for
  reference. Points of equal enthalpy (blue squares and red diamonds),
  obtained from constant pressure Monte Carlo simulations, 
give an
  estimate of the nanocrystal phase diagram as a function of shell
  thickness. The three phases are illustrated below the graph as cross
  sections of a $2$~nm shell crystal: wurtzite core and shell (WZ), a
  rocksalt core in a wurtzite shell (RS/WZ), and rocksalt core and
  shell (RS).}
\end{figure}

\begin{figure}
\includegraphics[width=0.9\textwidth]{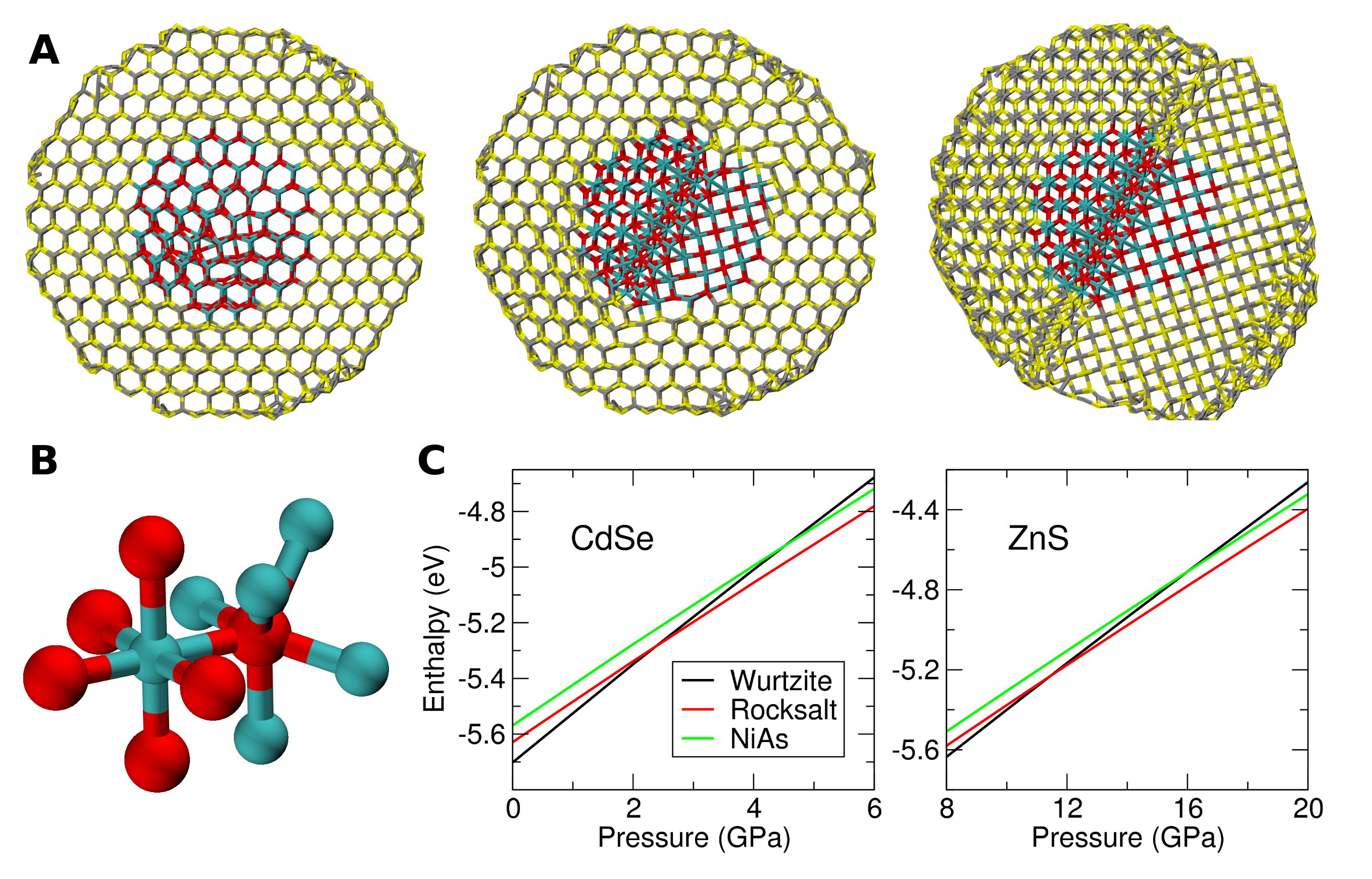}
\caption{\label{NiAs_fig} {\bf NiAs structure nucleates at high
    pressures.}  ({\bf A}) Time series of cross-sections 
of a $1.9$~nm-shell
  nanocrystal undergoing transforming from wurtzite to
  NiAs/rocksalt. The crystal is viewed along the wurtzite c-axis and
  the same set of atoms is displayed throughout. (Left) First stage of
  nucleation in the core at $17$~GPa.  (Center) $4$~ps later, no
  $4$-coordinated atoms remain in the core and a grain boundary
  between NiAs (upper left part of the crystal) and rocksalt (lower right part) 
is visible. The shell is
  visibly strained, but still wurtzite. (Right) At $20$~GPa, no
four-coordinated atoms remain;
the NiAs grain-boundary
  spans the entire crystal. ({\bf B}) Close-up view of a patch of CdSe
  in the NiAs structure, highlighting the different coordination
  environments of Cd (blue) and Se (red) atoms. ({\bf C}) Bulk
  enthalpies per atom as a function of pressure for CdSe and ZnS in the
  wurtzite, rocksalt and NiAs structures. Throughout the pressure
  range studied ($0$--$20$~GPa), NiAs is never stable. It is
  metastable with respect to the wurtzite structure at pressures
  larger than $4.5$~GPa and $16$~GPa, for CdSe and ZnS, respectively.}
\end{figure}

Points of equal enthalpy in Figure \ref{transp_fig} were determined in
NPT-Monte Carlo simulations as described in
Ref.~\cite{Grunwald2009a}. Initial configurations were extracted from
the molecular dynamics trajectories in which the transformations were observed. 
Bulk enthalpies in Figure
\ref{NiAs_fig} were calculated in NPT-Monte Carlo simulations of
approximately 500 
periodically replicated
atoms. Ewald sums were used for the calculation of
electrostatic interactions \cite{Frenkel2002}.

\begin{figure}
\includegraphics[width=\textwidth]{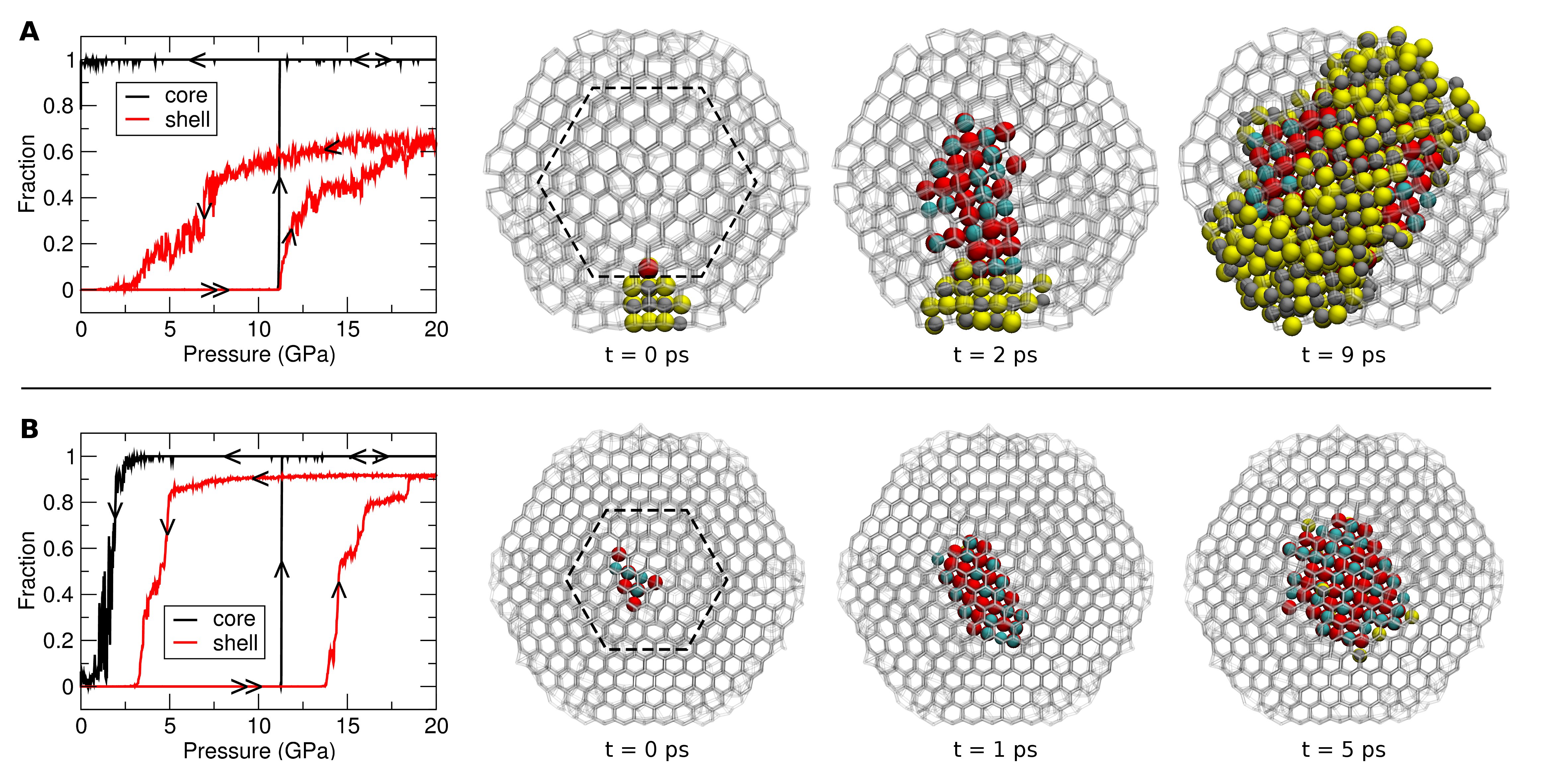}
\caption{\label{nucl_fig} {\bf The nucleation mechanism changes with
    increasing shell thickness.} ({\bf A}) (Left) Fraction of
  $6$-coordinated atoms as a function of pressure in the core (black)
  and shell (red) of a $1.2$~nm shell crystal. The transformation of
  both core and shell start around $11$~GPa. (Right) Cross sections
  highlight stages of the nucleation process, as seen along the
  wurtzite c-axis. The dashed line marks the interface between core and shell. Atoms that 
have undergone
a change of coordination are shown opaque. (For clarity, only clusters of 10 atoms or more are shown.)
 The transformation nucleates at the crystal
  surface and propagates inward. ({\bf B}) (Left) Fraction of
  $6$-coordinated atoms of a $2$~nm shell crystal. The transformations
  of the core and shell happen at different pressures, around $11$ and
  $14$~GPa, respectively. (Right) Snapshots show that the nucleus is
  located in the core.}
\end{figure}

\begin{figure}
\includegraphics[width=0.9\textwidth]{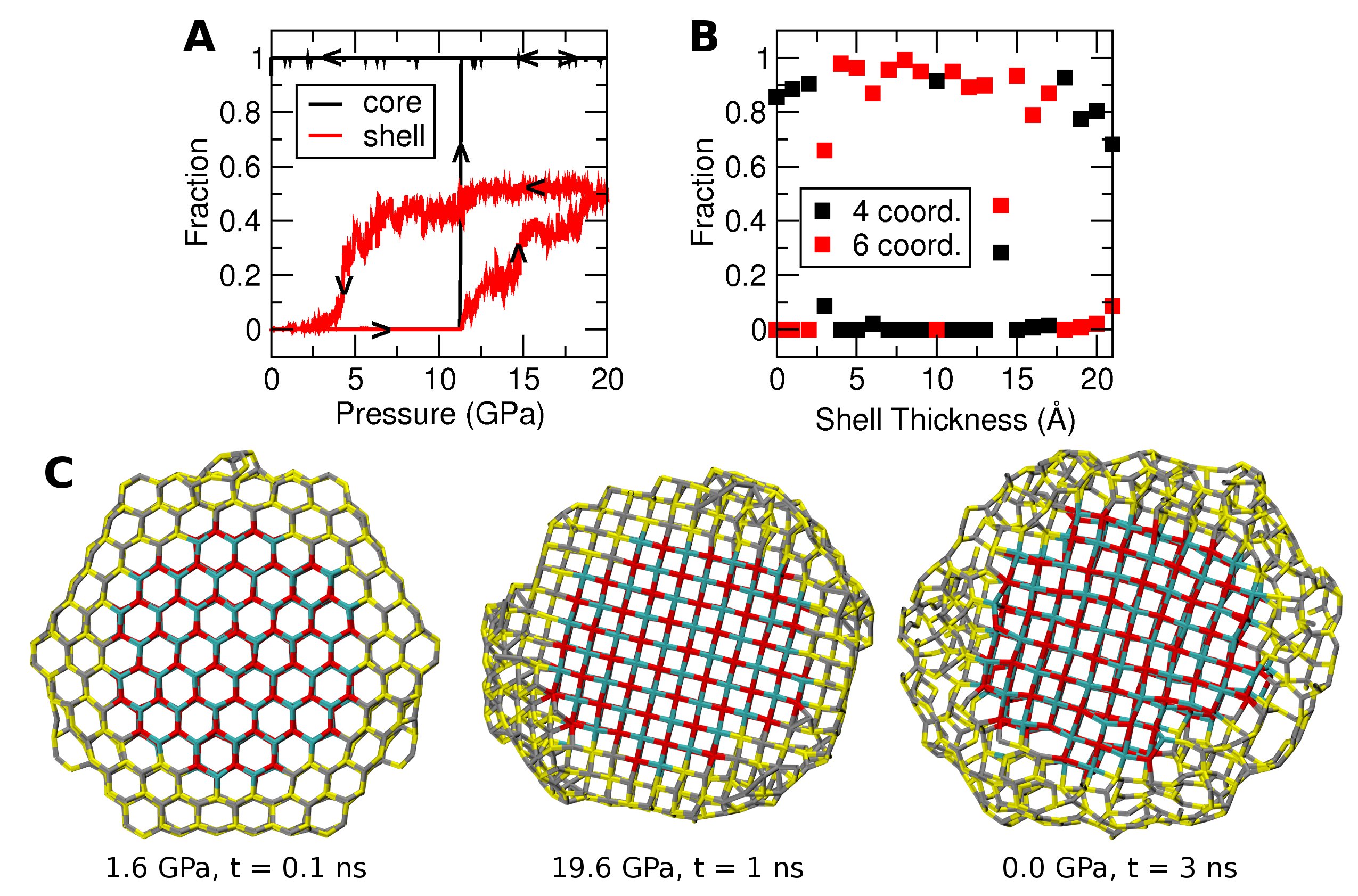}
\caption{\label{meta_fig} {\bf Rocksalt metastability at ambient
    pressure.} ({\bf A}) Fraction of $6$-coordinated atoms in the core
  (black) and shell (red) of a $0.9$~nm shell crystal as the pressure
  is increased to $20$~GPa and then released again. The rocksalt to
  wurtzite transformation at around $11$~GPa is well visible. While
  the shell undergoes the back-transformation at around $5$~GPa, the
  core remains in the rocksalt structure even at zero pressure. ({\bf
    B}) Fraction of $4$- and $6$-coordinated atoms in the core, $1$~ns
  after completing the pressure cycle. While crystals with very thin
  ($<0.4$~nm) and thick ($>1.7$~nm) shells transform back,
  crystals in a range of intermediate shell thicknesses remain in the
  rocksalt structure.  ({\bf C}) Cross sections of a $0.9$~nm shell
  crystal at different points in the pressure cycle, viewed along the
  wurtzite c-axis. (Left) At an up-stroke pressure of $1.6$~GPa, both
  core and shell are in the wurtzite structure. (Middle) At $19.6$~GPa
  the crystal is in the rocksalt structure.  (Right) A nanosecond
  after completing the pressure cycle, the rocksalt structure in the
  core persists. The shell has transformed back into a predominantly
  amorphous $4$-coordinated structure.}
\end{figure}

Hysteresis curves in Figures \ref{nucl_fig} and \ref{meta_fig} show
the fraction of 4- and 6-coordinated atoms in the core and shell,
respectively. These fractions were calculated as follows: For every
configuration along a given trajectory, the radial pair distribution
function $g(r)$ was calculated separately for core and shell atoms and
averaged over 5 ps intervals. The first minimum of $g(r)$ was found
numerically and the respective distance used as a cutoff for the
calculation of the number of nearest neighbors. To minimize 
interfacial contributions,
only core atoms within $10$~\AA{} of the center of mass were
included in the calculation; likewise, shell atoms within $18$~\AA{}
of the center were excluded.

\clearpage

\clearpage

\begin{thebibliography}{10}
\expandafter\ifx\csname url\endcsname\relax
  \def\url#1{\texttt{#1}}\fi
\expandafter\ifx\csname urlprefix\endcsname\relax\def\urlprefix{URL }\fi
\providecommand{\bibinfo}[2]{#2}
\providecommand{\eprint}[2][]{\url{#2}}

\bibitem{Alivisatos1994}
\bibinfo{author}{Tolbert, S.~H.} \& \bibinfo{author}{Alivisatos, A.~P.}
\newblock \bibinfo{title}{{Size Dependence of a First Order Solid-Solid Phase
  Transition: The Wurtzite to Rock Salt Transformation in CdSe Nanocrystals}}.
\newblock \emph{\bibinfo{journal}{Science}} \textbf{\bibinfo{volume}{265}},
  \bibinfo{pages}{373--376} (\bibinfo{year}{1994}).

\bibitem{Alivisatos1997}
\bibinfo{author}{Chen, C.-C.}, \bibinfo{author}{Herhold, A.~B.},
  \bibinfo{author}{Johnson, C.~S.} \& \bibinfo{author}{Alivisatos, A.~P.}
\newblock \bibinfo{title}{{Size Dependence of Structural Metastability in
  Semiconductor Nanocrystals}}.
\newblock \emph{\bibinfo{journal}{Science}} \textbf{\bibinfo{volume}{276}},
  \bibinfo{pages}{398} (\bibinfo{year}{1997}).

\bibitem{Alivisatos2001}
\bibinfo{author}{Jacobs, K.}, \bibinfo{author}{Zaziski, D.},
  \bibinfo{author}{Scher, E.~C.}, \bibinfo{author}{Herhold, A.~B.} \&
  \bibinfo{author}{Alivisatos, A.~P.}
\newblock \bibinfo{title}{{Activation Volumes for Solid-Solid Transformations
  in Nanocrystals}}.
\newblock \emph{\bibinfo{journal}{Science}} \textbf{\bibinfo{volume}{293}},
  \bibinfo{pages}{1803} (\bibinfo{year}{2001}).

\bibitem{Grunwald2009}
\bibinfo{author}{Gr\"{u}nwald, M.} \& \bibinfo{author}{Dellago, C.}
\newblock \bibinfo{title}{{Nucleation and growth in structural transformations
  of nanocrystals.}}
\newblock \emph{\bibinfo{journal}{Nano Lett.}} \textbf{\bibinfo{volume}{9}},
  \bibinfo{pages}{2099--102} (\bibinfo{year}{2009}).

\bibitem{Grunwald2006}
\bibinfo{author}{Gr\"{u}nwald, M.}, \bibinfo{author}{Rabani, E.} \&
  \bibinfo{author}{Dellago, C.}
\newblock \bibinfo{title}{{Mechanisms of the Wurtzite to Rocksalt
  Transformation in CdSe Nanocrystals}}.
\newblock \emph{\bibinfo{journal}{Phys. Rev. Lett.}}
  \textbf{\bibinfo{volume}{96}}, \bibinfo{pages}{255701}
  (\bibinfo{year}{2006}).

\bibitem{Farvid2009}
\bibinfo{author}{Farvid, S.~S.}, \bibinfo{author}{Dave, N.},
  \bibinfo{author}{Wang, T.} \& \bibinfo{author}{Radovanovic, P.~V.}
\newblock \bibinfo{title}{{Dopant-Induced Manipulation of the Growth and
  Structural Metastability of Colloidal Indium Oxide Nanocrystals}}.
\newblock \emph{\bibinfo{journal}{J. Phys. Chem. C}}
  \textbf{\bibinfo{volume}{113}}, \bibinfo{pages}{15928--15933}
  (\bibinfo{year}{2009}).

\bibitem{Alivisatos2002a}
\bibinfo{author}{Jacobs, K.}, \bibinfo{author}{Wickham, J.} \&
  \bibinfo{author}{Alivisatos, A.~P.}
\newblock \bibinfo{title}{{Threshold Size for Ambient Metastability of Rocksalt
  CdSe Nanocrystals}}.
\newblock \emph{\bibinfo{journal}{J. Phys. Chem. B}}
  \textbf{\bibinfo{volume}{106}}, \bibinfo{pages}{3759} (\bibinfo{year}{2002}).

\bibitem{Bawendi1999}
\bibinfo{author}{Dinega, D.~P.} \& \bibinfo{author}{Bawendi, M.~G.}
\newblock \bibinfo{title}{{A Solution-Phase Chemical Approach to a New Crystal
  Structure of Cobalt.}}
\newblock \emph{\bibinfo{journal}{Angewandte Chemie International Edition}}
  \textbf{\bibinfo{volume}{38}}, \bibinfo{pages}{1788--1791}
  (\bibinfo{year}{1999}).

\bibitem{Callen1985}
\bibinfo{author}{Callen, H.~B.}
\newblock \emph{\bibinfo{title}{{Thermodynamics and an Introduction to
  Thermostatistics}}} (\bibinfo{publisher}{John Wiley \& Sons},
  \bibinfo{address}{New York}, \bibinfo{year}{1985}), \bibinfo{edition}{2} edn.

\bibitem{Bealing2010a}
\bibinfo{author}{Bealing, C.}, \bibinfo{author}{Fugallo, G.},
  \bibinfo{author}{Martonak, R.} \& \bibinfo{author}{Molteni, C.}
\newblock \bibinfo{title}{Constant pressure molecular dynamics simulations for
  ellipsoidal{,} cylindrical and cuboidal nano-objects based on inertia tensor
  information}.
\newblock \emph{\bibinfo{journal}{Phys. Chem. Chem. Phys.}}
  \textbf{\bibinfo{volume}{12}}, \bibinfo{pages}{8542--8550}
  (\bibinfo{year}{2010}).

\bibitem{Alivisatos1995}
\bibinfo{author}{Tolbert, S.~H.} \& \bibinfo{author}{Alivisatos, A.~P.}
\newblock \bibinfo{title}{{The wurtzite to rock salt structural transformation
  in CdSe nanocrytsals under high pressure}}.
\newblock \emph{\bibinfo{journal}{J. Chem. Phys.}}
  \textbf{\bibinfo{volume}{102}}, \bibinfo{pages}{4642} (\bibinfo{year}{1995}).

\bibitem{Grunwald2009a}
\bibinfo{author}{Gr\"{u}nwald, M.} \& \bibinfo{author}{Dellago, C.}
\newblock \bibinfo{title}{{Transition state analysis of solid-solid
  transformations in nanocrystals.}}
\newblock \emph{\bibinfo{journal}{J. Chem. Phys.}}
  \textbf{\bibinfo{volume}{131}}, \bibinfo{pages}{164116}
  (\bibinfo{year}{2009}).

\bibitem{Spanhel1987}
\bibinfo{author}{Spanhel, L.}, \bibinfo{author}{Haase, M.},
  \bibinfo{author}{Weller, H.} \& \bibinfo{author}{Henglein, A.}
\newblock \bibinfo{title}{Photochemistry of colloidal semiconductors. 20.
  surface modification and stability of strong luminescing cds particles}.
\newblock \emph{\bibinfo{journal}{J. Am. Chem. Soc.}}
  \textbf{\bibinfo{volume}{109}}, \bibinfo{pages}{5649--5655}
  (\bibinfo{year}{1987}).

\bibitem{Hines1996}
\bibinfo{author}{Hines, M.~A.} \& \bibinfo{author}{Guyot-Sionnest, P.}
\newblock \bibinfo{title}{Synthesis and characterization of strongly
  luminescing zns-capped cdse nanocrystals}.
\newblock \emph{\bibinfo{journal}{J. Phys. Chem.}}
  \textbf{\bibinfo{volume}{100}}, \bibinfo{pages}{468} (\bibinfo{year}{1996}).

\bibitem{Bawendi1997}
\bibinfo{author}{Dabbousi, B.~O.} \emph{et~al.}
\newblock \bibinfo{title}{{( CdSe ) ZnS Core-Shell Quantum Dots : Synthesis and
  Characterization of a Size Series of Highly Luminescent Nanocrystallites}}.
\newblock \emph{\bibinfo{journal}{J. Phys. Chem. B}}
  \textbf{\bibinfo{volume}{101}}, \bibinfo{pages}{9463--9475}
  (\bibinfo{year}{1997}).

\bibitem{Peng1997}
\bibinfo{author}{Peng, X.}, \bibinfo{author}{Schlamp, M.~C.},
  \bibinfo{author}{Kadavanich, A.~V.} \& \bibinfo{author}{Alivisatos, A.~P.}
\newblock \bibinfo{title}{{Epitaxial Growth of Highly Luminescent CdSe/CdS
  Core/Shell Nanocrystals with Photostability and Electronic Accessibility}}.
\newblock \emph{\bibinfo{journal}{Journal of the American Chemical Society}}
  \textbf{\bibinfo{volume}{119}}, \bibinfo{pages}{7019--7029}
  (\bibinfo{year}{1997}).

\bibitem{Rosenthal2007}
\bibinfo{author}{Rosenthal, S.}, \bibinfo{author}{Mcbride, J.},
  \bibinfo{author}{Pennycook, S.} \& \bibinfo{author}{Feldman, L.}
\newblock \bibinfo{title}{{Synthesis, surface studies, composition and
  structural characterization of CdSe, core/shell and biologically active
  nanocrystals}}.
\newblock \emph{\bibinfo{journal}{Surface Science Reports}}
  \textbf{\bibinfo{volume}{62}}, \bibinfo{pages}{111--157}
  (\bibinfo{year}{2007}).

\bibitem{Li2011}
\bibinfo{author}{Li, Z.} \emph{et~al.}
\newblock \bibinfo{title}{{The structural transition behavior of CdSe/ZnS
  core/shell quantum dots under high pressure}}.
\newblock \emph{\bibinfo{journal}{Physica Status Solidi (B)}}
  \textbf{\bibinfo{volume}{248}}, \bibinfo{pages}{1149--1153}
  (\bibinfo{year}{2011}).

\bibitem{Rabani2001}
\bibinfo{author}{Rabani, E.}
\newblock \bibinfo{title}{{Structure and electrostatic properties of passivated
  CdSe nanocrystals}}.
\newblock \emph{\bibinfo{journal}{J. Chem. Phys.}}
  \textbf{\bibinfo{volume}{115}}, \bibinfo{pages}{1493} (\bibinfo{year}{2001}).

\bibitem{Rabani2002}
\bibinfo{author}{Rabani, E.}
\newblock \bibinfo{title}{{An interatomic pair potential for cadmium
  selenide}}.
\newblock \emph{\bibinfo{journal}{J. Chem. Phys.}}
  \textbf{\bibinfo{volume}{116}}, \bibinfo{pages}{258} (\bibinfo{year}{2002}).

\bibitem{Grunwald2006a}
\bibinfo{author}{Gr\"{u}nwald, M.} \& \bibinfo{author}{Dellago, C.}
\newblock \bibinfo{title}{{Ideal gas pressure bath: a method for applying
  hydrostatic pressure in the computer simulation of nanoparticles}}.
\newblock \emph{\bibinfo{journal}{Mol. Phys.}} \textbf{\bibinfo{volume}{104}},
  \bibinfo{pages}{3709} (\bibinfo{year}{2006}).

\bibitem{Grunwald2007}
\bibinfo{author}{Gr\"{u}nwald, M.}, \bibinfo{author}{Geissler, P.~L.} \&
  \bibinfo{author}{Dellago, C.}
\newblock \bibinfo{title}{{An efficient transition path sampling algorithm for
  nanoparticles under pressure}}.
\newblock \emph{\bibinfo{journal}{J. Chem. Phys.}}
  \textbf{\bibinfo{volume}{127}}, \bibinfo{pages}{154718}
  (\bibinfo{year}{2007}).

\bibitem{Wittenberg2009}
\bibinfo{author}{Wittenberg, J.}, \bibinfo{author}{Merkle, M.} \&
  \bibinfo{author}{Alivisatos, A. P.}
\newblock \bibinfo{title}{{Wurtzite to Rocksalt Phase Transformation of Cadmium
  Selenide Nanocrystals via Laser-Induced Shock Waves: Transition from Single
  to Multiple Nucleation}}.
\newblock \emph{\bibinfo{journal}{Phys. Rev. Lett.}}
  \textbf{\bibinfo{volume}{103}}, \bibinfo{pages}{1--4} (\bibinfo{year}{2009}).

\bibitem{Ithurria2007}
\bibinfo{author}{Ithurria, S.}, \bibinfo{author}{Guyot-Sionnest, P.},
  \bibinfo{author}{Mahler, B.} \& \bibinfo{author}{Dubertret, B.}
\newblock \bibinfo{title}{Mn$^{2+}$ as a radial pressure gauge in colloidal
  core/shell nanocrystals}.
\newblock \emph{\bibinfo{journal}{Phys. Rev. Lett.}}
  \textbf{\bibinfo{volume}{99}}, \bibinfo{pages}{265501}
  (\bibinfo{year}{2007}).

\bibitem{Fan2007}
\bibinfo{author}{Fan, H.~M.} \emph{et~al.}
\newblock \bibinfo{title}{{High pressure photoluminescence and Raman
  investigations of CdSe∕ZnS core/shell quantum dots}}.
\newblock \emph{\bibinfo{journal}{Appl. Phys. Lett.}}
  \textbf{\bibinfo{volume}{90}}, \bibinfo{pages}{021921}
  (\bibinfo{year}{2007}).

\bibitem{Frenkel2002}
\bibinfo{author}{Frenkel, D.} \& \bibinfo{author}{Smit, B.}
\newblock \emph{\bibinfo{title}{{Understanding Molecular Simulation}}}
  (\bibinfo{publisher}{Academic Press}, \bibinfo{address}{New York},
  \bibinfo{year}{2002}).

\end{thebibliography}
\end{document}